\documentclass[utf8]{FrontiersinVancouver} 

\setcitestyle{square} 
\usepackage{url,hyperref,lineno,microtype,subcaption}
\usepackage[onehalfspacing]{setspace}
\usepackage{amsmath}
\usepackage{amsfonts}
\usepackage{amssymb}
\graphicspath{{./figs/}}

\def\keyFont{\fontsize{8}{11}\helveticabold }
\def\firstAuthorLast{M\"unster {et~al.}} 
\def\Authors{Lambert Münster and Martin Weigel$^*$}
\def\Address{Institut für Physik, Technische Universität Chemnitz, 09107 Chemnitz, Germany}
\def\corrAuthor{Technische Universität Chemnitz, Institut für Physik, Reichenhainer Str.\ 70, 09126 Chemnitz, Germany}

\def\corrEmail{martin.weigel@physik.tu-chemnitz.de}
\def\pc{p_\mathrm{c}}

\begin{document}
\onecolumn
\firstpage{1}

\title[Spin glasses and percolation]{Spin glasses and percolation} 

\author[\firstAuthorLast ]{\Authors} 
\address{} 
\correspondance{} 

\extraAuth{}

\maketitle

\begin{abstract}

\section{}
The description of thermodynamic phase transitions in terms of percolation
transitions of suitably defined clusters has a long tradition and boasts a number of
important successes, the most prominent ones being in ferromagnetic lattice
models. Spin glasses and other frustrated systems are not among them as the clusters
of aligned spins usually considered in this context start to percolate in the
disordered phase and hence fail to indicate the onset of ordering. In this
mini-review we provide an overview of the state of the art in this field, including
recent advances, and outline the main open questions in the area.

\tiny
 \keyFont{ \section{Keywords:} spin glasses, percolation, lattice models, Monte Carlo
   simulations, phase transitions, quenched disorder} 
\end{abstract}

\section{Introduction}

Percolation models were first proposed and studied by Flory and Stockmeyer in the
context of polymer gelation \cite{flory:41,stockmayer:43}, and they have found
applications in an astonishingly broad range of areas, from forest fires
\cite{stauffer:book}, over porous media in oil fields \cite{king:01}, from electric
conductivity \cite{efros:76} all the way to complex networks \cite{cohen:00}. For
lattice systems, Broadbent and Hammersley \cite{broadbent:57} first proposed the idea
of what is today known as \emph{bond percolation}, where the edges of a lattice are
occupied at random with a probability $p$, and the resulting structure of connected
components is investigated \cite{grimmett:book}. This model provides one of the
simplest and most fundamental examples of a (usually continuous) phase
transition. For percolation, the transition is characterized by the appearance of a
spanning or \emph{incipient percolating} cluster that connects opposite edges of the
system and is of infinite size in the thermodynamic limit. At the transition point,
$\pc$, clusters on all length scales exist and the system is correlated up to the
largest distances, forming a (stochastic) self-similar fractal \cite{stauffer:book}.

This behavior is reminiscent of the spatial correlations observed in other systems
near criticality, for instance in the magnetic ordering transition of lattice spin
models \cite{mccoy:book}. In view of the success and intuitive appeal of the
percolation picture, it has been a longstanding goal in the description of phase
transitions and critical phenomena to view the ordering process in general systems as
a percolation transition of suitably defined structures or \emph{droplets} in the
substance undergoing an ordering process \cite{stanley:90}. Fisher proposed a model
\cite{fisher:67} that postulated droplets of a certain free energy whose average size
diverges at the critical point and that feature a cluster size distribution whose
exponents are related to the critical exponents of the thermal transition. A
microscopic definition of such droplets, however, was initially not available. While
it was clear that they must correspond to a spatially \emph{correlated} percolation
problem, it soon became clear that the clusters (connected components) of like spins
do \emph{not} fit Fisher's description as they percolate away from the thermal
critical point \cite{muller-krumbhaar:74}. Coniglio and Klein \cite{coniglio:80a}
first realized that suitable clusters resulted from a merely \emph{probabilistic}
occupation of bonds between like spins if the occupation probability was chosen as
$p=1-\exp(-2\beta J)$, where $\beta$ is the inverse temperature and $J$ denotes the
ferromagnetic exchange coupling. Independently, Fortuin and Kasteleyn
\cite{fortuin:72a} had provided a representation of the Potts model in form of a
correlated percolation model that contained the same cluster definition. The
resulting Fortuin-Kasteleyn-Coniglio-Klein (FKCK) clusters percolate at the thermal
transition point and their structure encodes the nature of spin-spin
correlations. They are also the basis for powerful numerical simulation schemes in
form of the cluster algorithms of Swendsen and Wang \cite{swendsen-wang:87a} as well
as Wolff \cite{wolff:89a}.

While these ideas are rather straightforwardly generalized from Ising to Potts
variables, as well as to continuous spins \cite{wolff:89a}, and even to disordered
ferromagnets \cite{ballesteros:98a}, they fail as soon as competing interactions and
frustration come into play \cite{kessler:90,arcangelis:91}. While FKCK clusters can
be easily generalized to this case by focusing on (parallel or antiparallel) spin
pairs with \emph{satisfied} bonds, it is found that in three dimensions such clusters
percolate far away from the spin-glass transition point as they, in fact, do
\emph{not} encode the relevant correlations at the spin-glass transition
\cite{fajen:20}. Instead, it has been proposed that one should consider cluster
definitions based on overlap variables, as they encode the order parameter of the
spin-glass transition \cite{machta:07}. Further, it appears that a more subtle
property of clusters than the mere onset of percolation might be associated with the
occurrence of the spin-glass transition. Only in two dimensions, where the situation
is somewhat different as the spin-glass transition is shifted to zero temperature,
one observes that for some types of overlap-based clusters the percolation points
asymptotically approach the spin-glass transition \cite{munster:23}.

Based on some of these observations, a number of cluster-update algorithms for spin
glasses have been proposed, the general target being to ensure that the updated
clusters undergo a percolation transition at or close to the spin-glass transition,
and that the structure of clusters encodes the correlations of the underlying spin
model. A general solution to this problem has not been found to date, but some
approaches provide reasonably good performance for systems in two dimensions
\cite{houdayer:01}, for spin glasses on diluted lattices \cite{joerg:05}, or for an
intermediate size range in three and higher dimensions \cite{zhu:15a}. In the
remainder of this mini-review, we will provide a more detailed discussion of the
connection between percolation and the spin-glass transition and the simulation
algorithms based on these observations.

\section{Spin clusters}

While some of what is discussed below can be generalized to the cases of Potts spins
as well as continuous models such as the \emph{XY} and Heisenberg spin glasses, to be
specific we focus on the case of the short-range (Edwards-Anderson) Ising spin glass
with Hamiltonian \cite{edwards:75a}
\begin{equation}
  {\mathcal H} = -\sum_{\langle i,j\rangle} J_{ij}\sigma_i \sigma_j,
\end{equation}
where $\sigma_i = \pm 1$ and the sum is taken over nearest-neighbor pairs of the
lattice only. To allow for a spin-glass phase, the distribution of the quenched
couplings $J_{ij}$ should include values of both signs, the most common cases being
the bimodal and Gaussian distributions. In a natural generalization from the cases of
ferromagnetic Ising and Potts models, FKCK clusters may be constructed for such a
system by occupying bonds between \emph{satisfied} spin pairs, i.e., those with
$J_{ij}\sigma_i \sigma_j > 0$, with probability
\begin{equation}
  p_\mathrm{FKCK} = 1-\exp[-2\beta|J_{ij}|].
  \label{eq:fkck-probability}
\end{equation}
There is clear numerical evidence that such clusters percolate at temperatures far
above the spin-glass transition, for instance at $T_\mathrm{c,FKCK} = 3.934(3)$ for
the three-dimensional symmetric $\pm J$ model \cite{arcangelis:91,fajen:20} as
compared to the spin-glass transition temperature at $T_\mathrm{SG} = 1.101(5)$
\cite{hasenbusch:08} (a similar difference is expected for the model with Gaussian
couplings). In two dimensions, these clusters percolate at
$T_\mathrm{SG} = 1.1894(3)$ for the Gaussian model \cite{munster:23}, while the
spin-glass transition only occurs for $T\to 0$ \cite{hartmann:01a}. More generally,
for a bimodal model with a fraction $x$ of antiferromagnetic bonds with $J_{ij} < 0$,
a coincidence of the percolation transition and the thermal transition point is only
observed for $x=0$ \cite{fajen:20}. This behavior is rather plausible since FKCK
clusters do not represent the relevant spin correlations in these systems. While for
the ferromagnet \cite{edwards:88a}
\begin{equation}
  \langle s_i s_j \rangle = \mathrm{Prob}(\text{$i$ and $j$ are connected by
    occupied bonds}),
    \label{eq:correlation_connectivity}
\end{equation}
the situation for spin glasses is more subtle, and one can show that in this case
\cite{coniglio:91,machta:07}
\begin{align}
    \langle s_i s_j \rangle &= \mathrm{Prob}(\text{$i$ and $j$ are connected by
                                an even number of antiferromagnetic bonds})\\
                              & -  \mathrm{Prob}(\text{$i$ and $j$ are connected by
                                an odd number of antiferromagnetic bonds}).
\end{align}
Hence, the percolation of FKCK clusters no longer implies the presence of long-range
order. Since the percolation transition of FKCK clusters does not encode spin-glass
criticality (but see Ref.~\cite{lundow:12} for a possible connection to damage
spreading), it is expected that it is in the universality class of random
percolation, and this expectation is borne out by the results of numerical simulation
studies \cite{imaoka:97,fajen:20,munster:23} as well as rigorous analysis
\cite{desantis:99,gandolfi:99}.

\section{Overlap clusters}

\begin{figure}
  \begin{center}
    \includegraphics[width=0.6\linewidth]{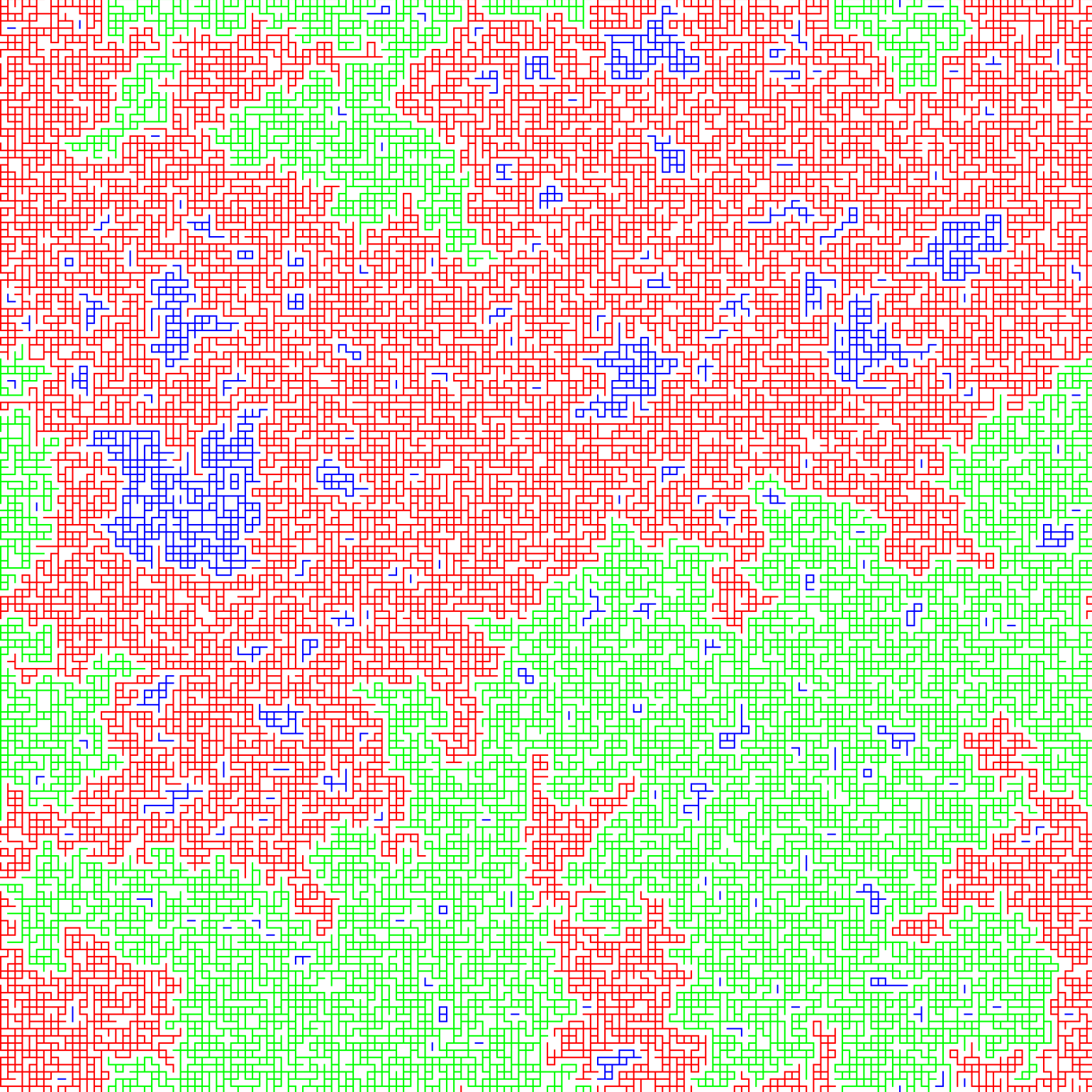}
    \caption{
      Typical configuration of CMR ``blue'' clusters in a sample of the 2D Gaussian
      Edwards-Anderson spin-glass model at low temperature ($\beta=3$). The red and
      green bonds correspond to the largest and second largest clusters, respectively.
    }
    \label{fig:snapshot}
  \end{center}
\end{figure}

\begin{figure}
  \begin{center}
    \includegraphics[width=0.6\linewidth]{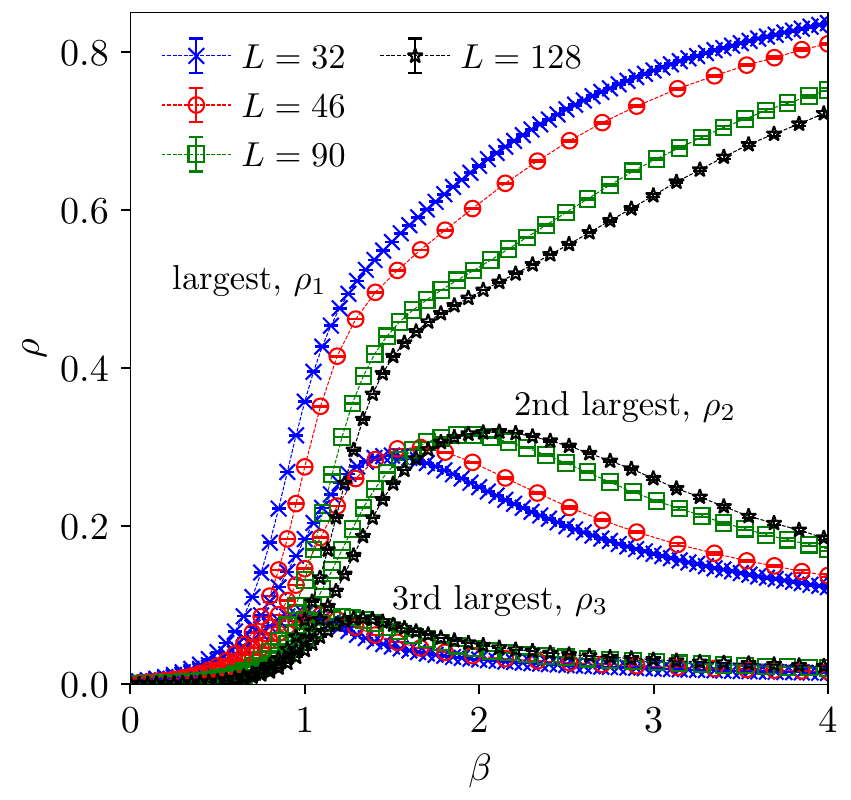}
    \caption{Average densities $\rho$ of the largest three CMR clusters in the 2D Gaussian
      Edwards-Anderson spin-glass model as a function of inverse temperature $\beta$
      for different lattice sizes $L$. At low temperatures, the combined weight of the
      two largest clusters increases with $L$.
    }
    \label{fig:blueclusters}
  \end{center}
\end{figure}

This failure is not surprising in view of the fact that the magnetization is no order
parameter for the spin-glass transition and, instead, for its description one needs to
turn to \emph{overlap} variables \cite{parisi:79}. Several cluster definitions have
been suggested based on the site or link overlap of two spin configurations using the
same disorder realization. Initially in the context of random-field models, Chayes,
Machta and Redner \cite{chayes:98} proposed a representation where doubly satisfied
(``blue'') bonds in a two-replica representation are occupied with a probability
\begin{equation}
  p_\mathrm{CMR,blue} = 1-\exp[-4\beta|J_{ij}|],
  \label{eq:blue-probability}
\end{equation}
while, additionally, singly satisfied (``red'') bonds are occupied with probability
\[
  p_\mathrm{CMR,red} = 1-\exp[-2\beta|J_{ij}|].
\]
Then, it is possible to relate the percolation properties of such clusters to the
occurrence of symmetry breaking of the spin system \cite{machta:07}: in contrast to
the ferromagnet, where the appearance of a percolating cluster suffices to indicate
the onset of the ordered phase, for spin glass there should be a ``blue'' cluster of
strictly larger density than any other cluster \cite{machta:09a}. In practise, one
observes the occurrence of \emph{two} percolating clusters of opposite overlap that
develop a density difference at the spin-glass transition
\cite{machta:07,munster:23}. A corresponding overlap configuration is shown in
Fig.~\ref{fig:snapshot} for the example of the 2D Gaussian spin glass, illustrating
that there are mainly two large clusters of opposite overlap, with any further
clusters being much smaller. If the weight of such smaller clusters diminishes for
systems of increasing sizes, the overlap $q$ (i.e., the order parameter) is connected
to the density difference of the two largest clusters. This is rigorously the case in
the mean-field Sherrington-Kirkpatrick model \cite{machta:07}, and numerical data in
3D \cite{machta:07} and 2D \cite{munster:23} are also consistent with such a picture
--- for the 2D case this is demonstrated by the data shown in
Fig.~\ref{fig:blueclusters} that consist of the densities of the three largest
clusters as a function of inverse temperature and for different lattice sizes. The
onset of percolation of CMR clusters itself again occurs away from the spin-glass
transition, with $T_\mathrm{c,CMR} \approx 3.85$ for the 3D bimodal model
\cite{machta:07} (which is surprisingly close to $T_\mathrm{c,FKCK}$); in 2D the CMR
percolation temperatures of finite lattices converge to $T=0$ for $L\to\infty$
\cite{munster:23}, consistent with the spin-glass transition temperature there.

Another possible cluster definition based on the overlap of two replicas results from
a simple duplication of the FKCK construction on the two spin configurations, i.e.,
bonds are occupied independently in the two replicas according to the FKCK
probability (\ref{eq:fkck-probability}), and clusters are constructed over all bonds
simultaneously occupied in \emph{both} replicas \cite{newman:07}. This is equivalent
to a bond occupation probability
\begin{equation}
  \label{eq:trfk-probability}
  p_\mathrm{TRFK} = (1-\exp[-2\beta|J_{ij}|])^2
\end{equation}
for doubly satisfied bonds. Such clusters might be referred to as two-replica
Fortuin-Kasteleyn (TRFK) clusters. These clusters behave rather similarly to the CMR
ones \cite{machta:07,munster:23}, which is not surprising as they follow the same
construction apart from the smaller bond occupation probability
$p_\mathrm{TRFK} \le p_\mathrm{CMR,blue}$. The latter leads to a significant
suppression of the percolation point which now occurs for
$T_\mathrm{c,TRFK} \approx 1.77$ \cite{machta:07}.

Finally, a cluster definition based on a \emph{site} percolation problem rather that
a \emph{bond} percolation one was first proposed in connection with a specific
cluster-update algorithm for spin glasses in 2D \cite{houdayer:01}. There, clusters
are grown in regions of constant overlap, and neighboring sites of the same overlap
are \emph{unconditionally} added to the cluster, such that the effective bond
occupation probability is
\[
  p_\mathrm{Houdayer} = 1.
\]
One might hence think of these as some form of \emph{geometric clusters} in overlap
space \cite{akritidis:23}. The CMR and TRFK clusters clearly are subregions of the
Houdayer clusters as for the latter one does not take into account whether a given
bond is satisfied or not. In many 3D lattices such as the simple cubic one, such
clusters percolate already for $T_\mathrm{c,Houdayer} = \infty$ as their site
percolation thresholds are $p_\mathrm{c} < 0.5$. Hence they have not been studied in
much detail there. In 2D, on the other hand, they again percolate at a sequence of
temperatures that approaches $T_\mathrm{SG} = 0$, but they are found to be in general
larger that the CMR and TRFK clusters \cite{munster:23}.

\section{Cluster updates}

In view of the spectacular success of cluster updates in alleviating critical slowing
down for ferromagnetic spin models \cite{swendsen-wang:87a,wolff:89a} it has been a
natural idea to use cluster moves to counter the dramatically slow dynamics observed in
spin-glass systems. In fact, the first proposal in this direction \cite{swendsen:86}
even (slightly) predates the ferromagnetic algorithms. Unfortunately, the cluster
component in this approach was not found to be extremely efficient, while the replica
component eventually lead to the development of the replica exchange or parallel
tempering method \cite{geyer:91,hukushima:96a} that is the \emph{de facto} standard for
spin-glass simulations.

Houdayer's proposal \cite{houdayer:01} for a cluster update for 2D models was in this
sense more successful. Geometric clusters are constructed in the way described above,
by connecting neighboring sites of equal overlap, and an update consists of flipping
the spins inside a cluster in \emph{both} replicas. Crucially, such updates can be
performed unconditionally, i.e., without adding an extra acceptance step, since they
leave the total energy of the replicated system invariant. Usage of more than one
replica (per temperature) is possible, but usually not found to be efficient
computationally \cite{kumar:20}. Due to the fixed energy, it is clear that such
updates are not ergodic, and hence need to be complemented, e.g., by single-spin flip
moves. While this approach works well on the square lattice, where the percolation
threshold $p_\mathrm{c} \approx 0.59 > \frac{1}{2}$ and the percolation points of
Houdayer clusters approach $T_\mathrm{SG} = 0$ for $L\to\infty$ \cite{munster:23},
the method is not very efficient in 3D, which is blamed on the fact that, for most
lattices in 3D, $p_\mathrm{c} < \frac{1}{2}$ \cite{houdayer:01}. In an attempt to
improve on this aspect, Zhu {\em et al.} \cite{zhu:15a} proposed a modification of
Houdayer's method where they grow a single cluster in the \emph{minority} phase of
the overlap, which is claimed to somewhat improve the performance in 3D. More
recently, the cluster selection for updates has been scrutinized in a multi-cluster
version of the algorithm discussed in the context of combinatorial optimization
problems \cite{vandenbroucque:22}, a close relative of spin-glass problems (see,
e.g., Ref.~\cite{perera:20}).

The CMR representation also suggests several cluster updates. Constructing only the
blue clusters, these can flip freely as the cluster construction rules together with
the bond occupation probability (\ref{eq:blue-probability}) mean that the update
satisfies detailed balance with respect to the equilibrium distribution
\cite{munster:23}. This was used by J\"org \cite{joerg:05} to efficiently simulate
spin glasses on diluted lattices, leading to overall smaller clusters. By
construction, however, the update is not ergodic since spins connected by (partially)
unsatisfied bonds cannot be updated. An extension proposed in Ref.~\cite{chayes:98}
(see also \cite{machta:07}) uses both red and blue bonds to construct blue and
\emph{grey} clusters, leading to a rejection-free and ergodic update which, however,
is still found to be relatively inefficient due to the onset of cluster percolation
above $T_\mathrm{SG}$ \cite{machta:07}.

\section{Discussion}

While a percolation perspective onto spin glasses and other frustrated systems has
not led to the same level of revolutionary success this approach has seen for
ferromagnets, significant progress has been possible. The cluster construction rules
used for ferromagnets (Fortuin-Kasteleyn--Coniglio-Klein), while applicable to spin
glasses, do not lead to structures that reflect spin-glass correlations. Instead,
clusters must be constructed in overlap space, corresponding to the order parameter
of the spin-glass transition. While there is no one-to-one correspondence between the
spin-glass transition and a simple percolation transition of a cluster type that has
been investigated to date, an intriguing picture has emerged: for the CMR and TRFK
clusters defined on two replicas \emph{two} equally large percolating clusters appear
significantly above the spin-glass transition and it is only at the spin-glass
transition that their densities start to differ \cite{machta:07,munster:23}. It
appears that below the percolation point smaller clusters beneath the two dominating
ones are asymptotically irrelevant.

Regarding cluster updates, a fundamentally efficient algorithm only exists in two
dimensions, while attempts for more general, and in particular, 3D systems have only
partially been successful. While some improved results where found in cases where the
average sizes of clusters constructed are reduced such as in diluted systems
\cite{joerg:05} or with algorithmic modifications \cite{zhu:15a}, it is not fully
clear whether such size reduction is a sufficient condition for improving
performance.

In view of this state of affairs a number of interesting questions remain to be
addressed in future studies. Is it possible to construct clusters that percolate at
or very close to the temperature of the spin-glass transition? One promising
direction in this respect is the study of \emph{multi-replica} overlaps
\cite{munster:23}. In view of Eq.~(\ref{eq:trfk-probability}) it is clear that,
depending on their precise construction, such clusters could percolate at lower and
lower temperatures as the number of replicas is increased. Regarding the algorithms,
it was seen that for blue clusters there are two very dominant large clusters in the
vicinity of the glass transition, such that in contrast to the ferromagnetic case
close to the transition there is no multi-scale nature of spin updates for such blue
clusters close to the spin-glass transition. This is likely the prime reason for the
unsatisfactory performance of such algorithms. In contrast, for Houdayer's algorithm
and its extensions, what is the cluster-size distribution? How do multi-cluster
variants of such algorithms perform as compared to the default single-cluster ones?
Answers to (some of) these questions hold the potential for significantly advancing
our understanding of the spin-glass transition while simultaneously facilitating much
improved efficiency in simulating spin-glass systems with the hope of answering some
more of the fundamental open questions of this field.

\section*{Conflict of Interest Statement}

The authors declare that the research was conducted in the absence of any commercial or financial relationships that could be construed as a potential conflict of interest.

\section*{Author Contributions}
LM and MW conducted the literature review and wrote the manuscript.

\bibliographystyle{Frontiers-Vancouver} 
\bibliography{paper}



\end{document}